# Surface location of alkaline-earth atom impurities on helium nanodroplets

Yanfei Ren and Vitaly V. Kresin

*Department of Physics and Astronomy, University of Southern California,*

*Los Angeles, California 90089-0484*

**Abstract**

There has been notable uncertainty regarding the degree of solvation of alkaline-earth atoms, especially Mg, in free $^4$He nanodroplets. We have measured the electron energy dependence of the ionization yield of picked-up atoms. There is a qualitative shape difference between the yield curves of species solvated in the middle of the droplet and species located in the surface region; this difference arises from the enhanced role played by the Penning ionization process in the latter case. The measurements demonstrate that Mg, Ca, Sr and Ba all reside at or near the droplet surface.

*Introduction.* When beams of helium nanodroplets are doped with picked-up atoms or molecules ("impurities"), the opening question is: where does the impurity locate, i.e., is it solvated inside the droplet or sitting at its surface? The great majority of dopants find it energetically favorable to sink into the middle, but alkali atoms reside in surface "dimples" [1,2] because the Pauli repulsion between their *s* valence electrons and those of helium suppresses the interatomic attractive force.

The alkaline earth atoms are an interesting case because they are close neighbors of the alkalis. The optical spectra of Ca and Sr attached to $He_n$ are strongly shifted from both gas-phase and bulk helium-solvated absorption lines [3]. This supports the picture of these atoms being in surface dimples which are, however, deeper than those for the alkali case [4]. A similar conclusion has been drawn for Ba [5,6].

For the lighter Mg atom, however, the situation is unsettled. An estimate of the likelihood of solvation is provided by the dimensionless parameter [7] $\lambda = 2^{-1/6} \rho \varepsilon r_{min}/\sigma$, where $\rho$ and $\sigma$ are the density and surface tension of liquid helium, $\varepsilon$ and $r_{min}$ are the depth and position of the minimum of the impurity-$^4$He interatomic potential. For $\lambda > 1.9$ the impurity is expected to be solvated, and for $\lambda < 1.9$ it is expected to locate on the surface. However, different pair potentials for the Mg-$^4$He interaction yield $\lambda$ parameters both somewhat above and somewhat below the critical value [8], preventing a definite conclusion. Similarly, a quantum Monte Carlo study of Mg interacting with $He_{n=2-50}$ clusters [9] showed that different pair potentials lead to markedly different Mg solubility properties. A density-functional treatment concluded that Mg becomes solvated [4], but a recent first-principles calculation [10] predicted that it resides in a "cave" near the surface instead. The shape of laser-induced fluorescence excitation spectra of Mg in $He_n$ [8] is similar to that of Mg atoms solvated in bulk liquid helium. This was interpreted as proof of full interior solvation, but it is also conceivable that atoms burrowed relatively deep into the subsurface layer would already yield a bulk-like spectrum.

In previous studies of Li [11] and Na [12] atoms and small clusters on He nanodroplets, we observed that there exists a straightforward signature of surface-vs.-volume location of impurities. It is provided by the form of the ionization yield curves (ion signal intensity as a function of electron bombardment energy). As discussed in the aforementioned references and will be illustrated below, the ionization channels for the two types of impurities have characteristic and qualitatively different shapes. In the present work, we apply this technique to a diagnosis of the location of alkaline earth impurities.

*Experiment.* A beam of $^4He_n$ droplets is produced by standard low-temperature supersonic expansion of pure gas at 40 bar stagnation pressure through a 5 μm nozzle maintained at 12 K. These conditions were fixed for all the measurements reported here and correspond to an average droplet size of approximately $10^4$ He atoms [13].

The beam entered the pick-up chamber though a 0.4 mm skimmer, and passed though a rotating wheel chopper followed by a pick-up cell. Mg, Ca, Sr or Ba were loaded into the cell and heated up respectively to 230°C, 350°C, 365°C and 438°C, corresponding to vapor pressures in the range of $10^{-7}$ to $10^{-6}$ torr. These conditions resulted in the droplets picking up mostly single metal atoms (monomers gave the strongest signal in the metal ion mass spectrum). Reference ion yield curves were taken with xenon atom dopants, in which case the pick-up cell was fed from a lecture bottle of compressed gas through a length of 0.75 mm inner diameter tubing. To optimize the pick-up of single Xe atoms the bottle pressure regulator was set to 0.6 bar.

In the last chamber the doped droplets were detected by a quadrupole mass spectrometer (Balzers QMG-511). Feeding the mass spectrometer output into a lock-in amplifier synchronized with the beam chopper provided discrimination against background residual gas ions. Ion yield curves were determined by plotting the area of the impurity peak in the mass spectrum as the ionizing electron energy was varied from 18 eV to 60 eV at constant ionizer current.

***Results and discussion.*** Figs. 1 and 2 show the yield curves of $Ca^+$, $Sr^+$, $Ba^+$ and $Mg^+$, respectively, as well as of $Xe^+$. The data were derived from the peaks of pure monomer ions [14]. We observe that all metal ion yield curves are remarkably different from that of $Xe^+$. In fact, the former are of the same shape as for Li and Na impurities, while xenon is similar to $He_k^+$, $NaI^+$, $SF_6^+$ [11,12,15], etc. This qualitative contrast immediately implies that whereas Xe is fully solvated (cf., e.g., [16]), all of the investigated alkaline earth atoms are located near the droplet surface.

The character of the yield curves has been rationalized [11,12] by reference to the two relevant impurity ionization mechanisms [15]. The incoming electrons dominantly collide with a surface helium atom, as there are many of those and only one impurity. The outcome is either a positive "hole" $He^+$ or a metastable $He^*$ atom with one electron in the 2*s* state. For free helium atoms, the respective excitation thresholds are 22.4 eV and 19.8 eV, respectively. What is even more significant is that the shapes of the electron-impact excitation cross sections for these two products are very different, see Fig. 3.

The positive hole will very quickly [15] diffuse towards the center of the droplet, as is energetically favorable. There it can ionize a solvated impurity by charge transfer. Neutral $He^*$, on the other hand, finds no such gain from solvation. On the contrary, its excited electron creates a bubble state which (like alkali atoms) prefers to remain near the surface. Thus surface, rather than solvated, impurities are much more likely to encounter $He^*$ and to be ionized by the Penning mechanism (e.g., $He^*+X \rightarrow X^+ + He + e^-$ or $He^* + X \rightarrow XHe^+ + e$; note also that the cryogenic droplet environment implies very low $He^*+X$ collision energies which strongly enhances the reaction cross section [20]). As a result, the ionization yield curves for surface-resident impurities will reflect a strong contribution of the $He^*$ excitation shape (Fig. 3a).

In summary, a picture consistent with our electron-impact ionization data as well as with the previous spectroscopic results [3,5,8] is that the alkaline earth atomic dopants Ba, Sr, Ca, and Mg all reside in the surface region of helium nanodroplets, with magnesium sitting in a deeper burrow than the others.

***Acknowledgements.*** We would like to thank Prof. Marius Lewerenz for very useful discussions,

and Vincent Kan for assistance with experiments and data analysis. This work was supported by the U.S National Science Foundation under grant no. PHY-0245102.

# Figure captions

**Fig.1.** Ion yield curves for Ca, Sr, Ba, and Xe atoms picked up by a helium nanodroplet beam. (Lines connect data points to guide the eye.) The shape difference reflects the fact that the metal atoms are located at the droplet surface and the xenon atom is solvated inside the droplet.

**Fig.2**. Ionization yield curves for picked-up atoms of Mg and Xe (as in Fig. 1), indicating the difference in their locations.

**Fig. 3**. Cross sections for metastable excitation [17,18] (see [11] for details) and for electron-impact ionization [19] of the He atom.

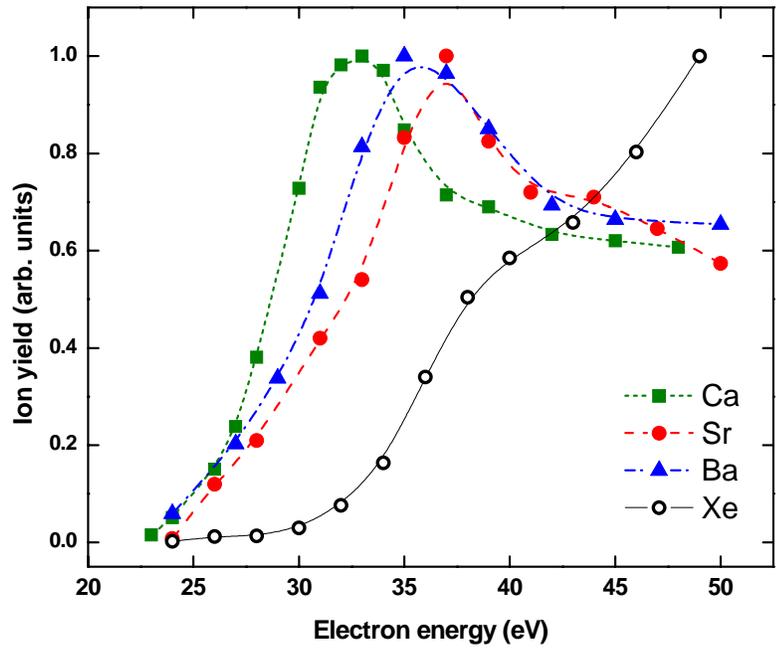

Fig.1

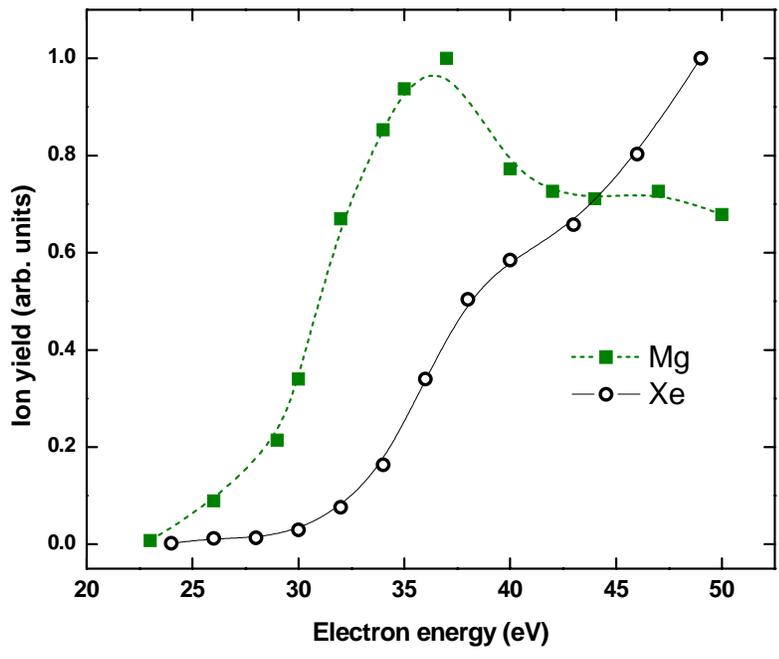

Fig.2

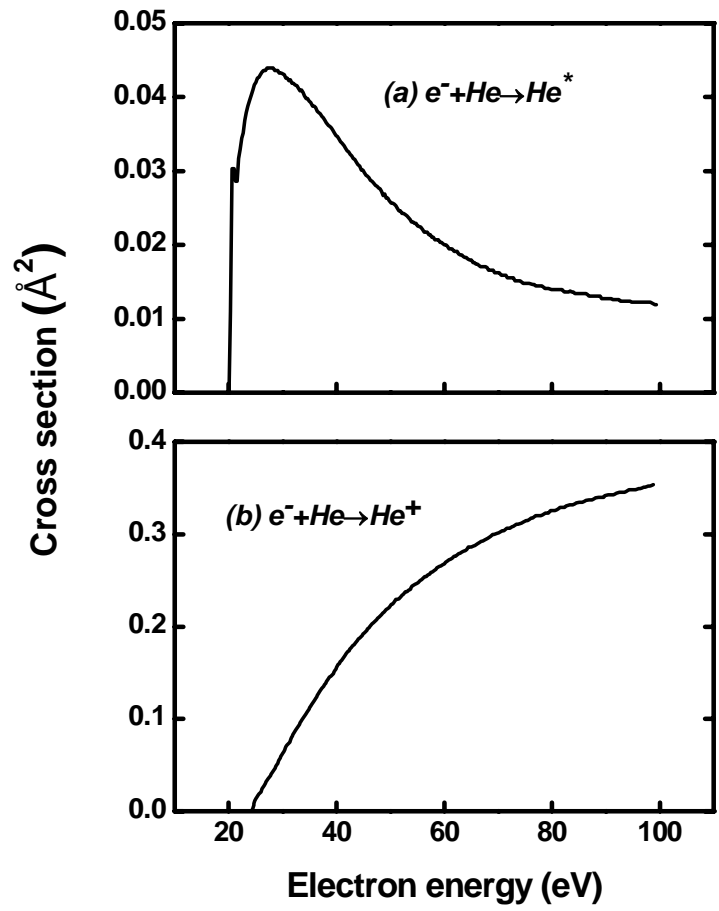

Fig. 3